# Effective phase diffusion for spin phase evolution under random nonlinear magnetic field

Guoxing Lin*

*Carlson School of Chemistry and Biochemistry, Clark University, Worcester, MA 01610, USA*

*Email: glin@clarku.edu

**ABSTRACT**

The general theoretical description of spin self-diffusion under nonlinear gradient is proposed, which extends the effective phase diffusion method for linear gradient field. Based on the phase diffusion, the proposed method reveals the general features of phase evolutions in non-nonlinear gradient fields. There are three types of phase evolutions: phase diffusion, float phase evolution, and shift based on the starting position. For spin diffusion near the origin of the nonlinear field, these three phase evolutions significantly affect the NMR signal. The traditional methods have difficulties in handling these three-phase evolutions. Notably, the phase from float phase evolution is missed or misplaced in traditional methods, which leads to incorrect NMR signal attenuation or phase shift. The method here shows that the diffusing and float phase evolutions come from the gradient field's first and second derivatives. Based on these three phase evolutions, the phase variance and corresponding NMR signal attenuation are obtained, demonstrated by calculating the phase diffusions under the parabolic and cubic fields. The results indicate that signal attenuation obeys Gaussian attenuation for a short time, then changes to Lorentzian or Mittag-Leffler function attenuations when time increases, significantly different from Gaussian attenuation. For spins starting diffusion far away from the origin, the signal attenuation is Gaussian, but the float phase still has an important effect on the total phase shift of even-order gradient fields, which could be used to measure the diffusion coefficient directly. Random walk simulations are performed, which support the obtained theoretical results. The general theoretical expressions are obtained, which can handle random order nonlinear gradient field. The results could help develop advanced experimental techniques based on a nonlinear gradient field in NMR and MRI.

## 1. INTRODUCTION

Employing magnetic field gradient to measure diffusion can be tracked back to Hahn, who first observed the influence of molecule diffusion under a gradient field upon echo amplitudes in 1950 [1]. Pulsed-field gradient (PFG) technique has broad applications in nuclear magnetic resonance (NMR) and magnetic resonance imaging (MRI) [2,3,4,5,6]. Under the gradient field, spins precess with position-dependent frequencies, and thus, their accumulated phases vary in different locations, which enables the gradient field to encode the spatial information for MRI experiments. Additionally, because different spin quantum coherences evolve at different frequencies in the gradient field, employing appropriate dephasing and rephasing gradient pulses' combination can selectively refocus the desired coherence pathway, which makes the gradient field an essential tool for selecting the coherence transfer pathway in modern NMR experiments. Furthermore, diffusing spins randomly move around, and their phase evolutions depend on their diffusion paths. Due to diffusion, the phase spreading of the spin system cannot be refocused even when dephasing and refocusing gradient pulses are employed. The phase spreading results in signal attenuation in NMR and MRI experiments [5,6,7]. Different spin phase distribution corresponds to different types of signal attenuations. The diffusion coefficient can be extracted by analyzing signal-intensity data in PFG diffusion experiments. The differences in diffusion parameters can be used as important contrast factors to build imaging for clinical studies such as acute stroke [8]. The accuracy of the analysis of PFG diffusion experimental data relies on the appropriate theoretical expressions [5,6,7].

Nonlinear gradient field encounters in many situations [9, 10,11,12, 13]. The frequently encountered PFG experiments employ linear gradient fields. However, a linear gradient field is an ideal situation. In actual samples, the internal local magnetic fields in many systems, such as porous material, are often inhomogeneous. Additionally, the imperfection of the external field, the eddy current, and the sample susceptibilities and shape result in inhomogeneous magnetic fields. These inhomogeneous fields are often nonlinear. The magnetic field inhomogeneity can significantly affect experimental results. It can result in artifacts in diffusion-based imaging and poor spatial resolution in MRI [12,14], and complicates the interpretation of NMR relaxation and diffusion results [15, 16]. Nonlinear gradient theories are necessary to interpret the effects of the nonlinear gradient field on experiments.

Although many theoretical efforts have been made to investigate the nonlinear gradient field [9,10,11,12,17,18], due to the complexity of the nonlinear field, the currently available theoretical results are incredibly insufficient. First, there are some apparent discrepancies in existing theoretical results. Even in the simplest nonlinear field, the parabolic field, the signal attenuations, and corresponding phase variances reported in the literature for spins starting diffusion from the origin of the parabolic field are different. Under a constant parabolic gradient field, Ref. [12] gives a signal attenuation $\exp\left(-\frac{7}{6}(\gamma g_2)^2 D^2 \delta^4\right)$ corresponding to a phase variance $\frac{7}{3}(\gamma g_2)^2 D^2 \delta^4$, where $\gamma$ is the gyromagnetic ratio, $g_2$ is the gradient field coefficient, $D$ is the diffusion coefficient, and $\delta$ is the gradient pulse length. In contrast, the Green function method gives a signal attenuation $\exp\left(-\frac{2}{3}(\gamma g_2)^2 D^2 \delta^4\right)$ [10], corresponding to a phase variance $\frac{4}{3}(\gamma g_2)^2 D^2 \delta^4$. While for two gradient pulses experiments, Refs. [10] does not give the free diffusion results but yields a signal attenuation $\exp\left(-\frac{4}{3}(\gamma g_2)^2 R_p^2 D \delta^3\right)$, corresponding to a phase variance $\frac{8}{3}(\gamma g_2)^2 R_p^2 D \delta^3$ for a short time diffusion restricted inside a pore-size $R_p$; when $R_p$ is infinite, the phase variance, however, cannot be reduced to $\frac{44}{3}(\gamma g_2)^2 D^2 \delta^4$ for free diffusion presented in Ref. [12].

Secondly, many of the theoretical results have obvious limits. Refs. [10] and [12] only treat the simplest nonlinear field, $n = 2$, the parabolic field. Additionally, the results in Ref. [12] are limited to diffusion starting from the origin. Meanwhile, diffusion near the origin is a challenge to many other methods. In non-origin position $z_0$, the time-dependent position $z(t)$ can be approximately treated as $z_0$ when $z_0 \gg 2Dt$. However, such an approximation for $z(t)$ is not appropriate for diffusion near the origin. Meanwhile, Refs. [10,18] do not directly treat the free diffusion under the parabolic field for the typical two pulses' sequences but extract the free diffusion results from restricted diffusion by setting $R_p$ as infinite, considering the discrepancies mentioned in the previous paragraph, directly obtaining the theoretical free diffusion result under two pulses is desirable.

Thirdly, there are some obvious challenges in the reported results. In Ref. [12], the attenuation expression from two pulses with a delay time cannot be reduced to the result of two pulses without delay. Additionally, in Ref. [10], an extra oscillatory phase term $g_2 D t^2$ is presented, but other methods do not have such a term. R. [10] hinted that such an oscillatory term results from the rapidly varying gradient field, which is unclear, particularly for the two-pulse experiment with a delay. The gradient field is off during the delay between pulses. This oscillatory term will be clearly obtained in the theoretical derivation in this paper, and a much clearer physical picture can be obtained.

Considering the difficulties of existing theories, it is necessary to develop new theoretical treatments to understand the nonlinear gradient field better. This paper proposes a versatile theoretical method for spin diffusion under the nonlinear field. The order of the gradient field can be any integer or non-integer, namely a real number or even a fraction. Additionally, the field can be a multiple-order gradient field, and the gradient pulse shape can be random rather than rectangular.

This method extends the effective phase diffusion equation method developed in Ref. [7]. Based on random walk theories, the phase diffusion approach is a powerful theoretical tool for spin dynamics, as spin phase evolution is a natural phase random walk process in many NMR phenomena. Phase diffusion-based methods have recently been applied to describe the phase evolution of spin coherence affected by the linear PFG [7,19], NMR relaxation [20], and NMR chemical exchange [21]. Compared to traditional NMR theories, phase diffusion-based methods possess certain advantages. It directly handles the phase evolution process in phase space, and the phase distribution can often be obtained. For example, a widely used PFG approximation for normal diffusion is the Gaussian phase distribution (GPD) approximation [5,6]; in contrast, for normal diffusion, the GPD is an exact solution from the phase diffusion equation [7]. Additionally, directly handling the phase evolution process in phase space often reduces the degree of solving complexity for analyzing NMR phenomena. Therefore, the phase diffusion methods can be straightforwardly applied to solve anomalous diffusions in PFG experiments [7], anomalous exchange processes [21], and fractional NMR relaxation [20], which are challenges to conventional theories. Furthermore, directly handling the phase evolution process in phase space can help reveal features that the traditional real space methods cannot find. In Ref. [22], the phase and time coupling are handled by coupled phase random walk, which gives a striking coupling constant in the spectral density expression for NMR relaxation. In Ref. [21], the phase diffusion method found that when the exchange time follows a distribution, the exchange time constant is twice as fast as the traditional result based on a single exchange time.

This paper's result clearly shows three types of phase evolutions in nonlinear gradient field experiments: phase diffusion, float phase evolution, and shift based on the starting position. Except for the shift phase evolution, the traditional theories do not have a clear concept of the other two types of phase evolutions, although the green function method indicates an extra oscillatory term in the magnetization. Ref. [10] explains that the oscillatory term results from rapidly varying magnetic field and magnetization. The method here clearly shows that the diffusing phase evolution comes from the first derivative of the gradient field, while the float phase evolution comes from the second derivative. The extra oscillatory term for the parabolic field comes from the accumulating phase from the float phase evolution. It affects magnetization regardless of whether there is a delay between gradient pulses. The float phase evolution can also be either a drift motion (in the parabolic field) or a diffusion (in the non-parabolic field).

Based on the effective phase diffusion, the phase diffusion coefficient, phase variance, and corresponding NMR signal attenuation can be calculated, which is demonstrated by calculating the phase diffusions under the parabolic and cubic fields. The signal attenuation obeys Gaussian attenuation for a short time, then changes to Lorentzian or Mittag-Leffler function (MLF) attenuations when time increases, significantly different from the Gaussian attenuation. The non-Gaussian attenuations come from Lorentzian or long-tailed phase distributions. The results of this paper could help us understand the cause of the discrepancy in existing theoretical results and provide many theoretical expressions that are not available from reported theories. The nonlinear gradient has some properties that are not available by the linear gradient field [23]. For diffusion near the origin of the nonlinear gradient field, the higher the order of the nonlinear gradient field, the higher order of diffusion coefficients and gradient pulse duration the signal attenuation depends on, which has advantages over the linear gradient field because it is relatively sensitive to the change of diffusion coefficient and pulse lengths. Hence, the theoretical results can potentially help develop advanced experimental techniques based on a nonlinear gradient field.

**2. THEORY**

2.1 phase random walk under a general nonlinear field

We first consider the one-dimensional anomalous diffusion under the magnetic field with the nonlinear field $B(z,t) = B_0 + B_g(z)$, $B_g(z) = \sum_{n\geq 1} g_n(t) \cdot f_n(z)$, where $B_0$ is the exterior magnetic field, $z$ is the position, and $g_n(t)$ is the coefficient of nonlinear field [5-7]. $\sum_{n\geq 1} g_n(t) \cdot f_n(z)$ rather than $gf(z)$ is employed because its derivation results are more general.

The magnetic field exerts a torque on each spin moment. The torque changes the spin angular momentum direction, which leads the spin to precess about the magnetic field with Larmor frequency $\omega(z,t) = \gamma B(z,t)$. In a rotating frame with angular frequency $\omega_0 = \gamma B_0$, a diffusing spin at position $z(t)$ has a time-dependent angular frequency $\gamma g(t) \cdot B_g(z(t))$, and its phase accumulated along the diffusion path is [7]

$$\phi(t) = \int_0^t \gamma B_g(z(t'))dt' = \int_0^t \gamma \sum_{n\geq 1} g_n(t') \cdot f_n(z(t')) \, dt', \tag{1}$$

where $\phi(t)$ is the net-accumulated phase. The range of $\phi(t)$ is $-\infty < \phi(t) < \infty$ rather than $-\pi \leq \phi(t) \leq \pi$, and $\cos(\phi(t))$ is the projection factor of the spin magnetization to the observing coordinate axis. Assuming the initial signal $S(0) = 1$, the NMR signal comes from the ensemble contribution from all spins by averaging over all possible phases [5-7]:

$$S(t) = \int_{-\infty}^{\infty} P(\phi,t) \exp(+i\phi) d\phi, \tag{2}$$

where $S(t)$ is the signal intensity at time $t$, and $P(\phi,t)$ is the accumulating phase probability distribution function. In a symmetric diffusion system, Equation (2) can be further written as $S(t) = \int_{-\infty}^{\infty} P(\phi,t) \cos(\phi) d\phi$. For simplicity, only diffusion with a symmetric probability distribution is studied here. The self-diffusion process can be described by a random walk, which consists of a sequence of independent random jumps with waiting times $\Delta t_1$, $\Delta t_2$, $\Delta t_3$,..., $\Delta t_m$, and corresponding displacement lengths $\Delta z_1$, $\Delta z_2$, $\Delta z_3$,..., $\Delta z_m$. Thus, $t_j = \sum_{i=1}^{j} \Delta t_i$, $z(t_j) = z_0 + \sum_{i=1}^{j} \Delta z_i$. Based on the random walk, Equation (1) can be rewritten as [7]

$$\phi(t) = \sum_{i=1}^{m} \gamma \Delta t_i B_g(z(t_i)) = \sum_{n\geq 1} \sum_{i=1}^{m} \gamma \Delta t_i g_n(t_i) \cdot f_n(z(t_i))$$
$$= \sum_{n\geq 1} \sum_{i=1}^{m} \gamma \Delta t_i g_n(t_i) \cdot \left(\sum_{j=1}^{i} \Delta h_j + h_0\right), \tag{3a}$$

where

$$\begin{cases} f_n(z(t_i)) = \sum_{j=1}^{i} \Delta h_j + h_0, \\ h_0 = f_n(z_0), \\ \Delta h_j \approx f_n'(z(t_{j-1})) \Delta z_j + \frac{1}{2} f_n''(z(t_{j-1}))(\Delta z_j)^2 + \cdots + \frac{1}{m!} f_n^m(z(t_{j-1}))(\Delta z_j)^m + \cdots, \end{cases} \tag{3b}$$

where Taylor expansion is used for obtaining $\Delta h_j$. Usually, the jump length is small and $(\Delta z_j)^2$ is much smaller than $|\Delta z_j|$, however, the accumulated effect of $(\Delta z_j)^2$ in the random walk path can often be comparable to that of $\Delta z_j$. For the higher orders, $(\Delta z_j)^m$, $m \geq 3$, which is negligible in both the single step and the accumulated effect. Therefore, only $\Delta h_j \approx f_n'(z(t_{j-1})) \Delta z_j + \frac{1}{2} f_n''(z(t_{j-1}))(\Delta z_j)^2$ will be considered here.

At time $t_{tot}$, the end of the gradient pulse, by interchange the order of the summarization of the time and space as proposed in ref [7], we have

$$\phi(t_{tot}) = \sum_{n\geq 1} \left[\sum_{j=1}^{m} (K_n(t_{tot}) - K_n(t_{j-1})) \Delta h_j + K_n(t_{tot}) h_0\right] \tag{4}$$

where

$$K_n(t) = \int_0^t \gamma g_n(t')dt'. \tag{5}$$

When $n = 1$, $K_1(t)$ is the wavenumber for the linear gradient field [5-7]. Substituted Eq. (3b) into Eq. (4), we have

$$\phi(t_{tot}) = \phi_D(t_{tot}) + \phi_{float}(t_{tot}) + \phi_{shift,z_0}(t_{tot})$$

$$= \sum_{j=1}^{m} \left[ \sum_{n \geq 1} \left( K_n(t_{tot}) - K_n(t_j) \right) f_n'\left(z(t_j)\right) \right] \Delta z_j + \frac{1}{2} \sum_{j=1}^{m} \left[ \sum_{n \geq 1} \left( K_n(t_{tot}) - K_n(t_j) \right) f_n''\left(z(t_j)\right) \right] (\Delta z_j)^2$$

$$+ \sum_{n \geq 1} K_n(t_{tot}) f_n(z_0).$$

(6a)

where $\phi_D(t_{tot})$, $\phi_{float}(t_{tot})$, and $\phi_{shift,z_0}(t_{tot})$ correspond to the first, second, and third term in the last line of Eq. (6a). Note that $t_{j-1}$ is replaced with $t_j$ in Eq. (6a) because their difference is negligible. $f_n'(z(t))$ and $f_n''(z(t))$ are correlated based on the same path of a particle when $n \neq 2$.

In Eq. (6a), the first term $\phi_D(t_{tot})$ is an obvious random phase walk as it is modified by the random $\Delta z_j$, while the second term, $\phi_{float}(t_{tot})$ could either be a drift-diffusion if $f_n''(z(t_j))$ is a constant as in the parabolic field case, or a random walk if $f_n''(z(t_j))$ is a $z(t_j)$-dependent function as in the cubic field case, which could be solved by effective phase diffusion presented in [7] or the method developed in this paper. The term "float" is used here because compared to the first term, the random walk steps in the second term keep the jump direction for a relatively long time, which is a sense of drift motion. The third term $\phi_{shift,z_0}(t_{tot})$ in Eq. (6a) for $K(t_{tot}) \neq 0$ is a phase shift depending on $(z_0)^n$. When the spins start the diffusion from the origin, $z_0 = 0$, $\phi_{shift,z_0}(t_{tot})$ equals to zero.

The practical pulse gradient experiments often employ both dephasing and refocusing gradient pulses, where $K(t_{tot}) = 0$. When $K(t_{tot}) = 0$, Eq. (6a) reduces to

$$\phi(t_{tot}) = \phi_D(t_{tot}) + \phi_{float}(t_{tot}) = -\sum_{j=1}^{m} \left[ \sum_{n \geq 1} K_n(t_j) f_n'\left(z(t_j)\right) \right] \Delta z_j -$$
$$\frac{1}{2} \sum_{j=1}^{m} \left[ \sum_{n \geq 1} K_n(t_j) f_n''\left(z(t_j)\right) \right] (\Delta z_j)^2 .$$

(6b)

When $f_n(z) = z$, $f_n''(z(t_j)) = 0$, and Eqs. (6a-b) reproduce the results for the linear gradient field presented in Ref. [7].

The effective phase diffusion coefficient for the first term $\phi_D(t_{tot})$ in Eqs. (6a-b) is

$$D_\phi(t) = \left[ \sum_{n \geq 1} (K_n(t_{tot}) - K_n(t)) f_n'(z(t)) \right]^2 \frac{\langle (\Delta z(t))^2 \rangle}{2\Delta t} = \left[ \sum_{n \geq 1} (K_n(t_{tot}) - K_n(t)) f_n'(z(t)) \right]^2 D.$$

(7)

By averaging over all possible $z(t)$ of the spin system, we have

$$D_\phi(t) = \left\langle \left[ \sum_{n \geq 1} (K_n(t_{tot}) - K_n(t)) f_n'(z(t)) \right]^2 \right\rangle D,$$

(8a)

where

$$\left\langle \left[ \sum_{n \geq 1} (K_n(t_{tot}) - K_n(t)) f_n'(z(t)) \right]^2 \right\rangle = \int_{-\infty}^{\infty} P(z, t) \left[ \sum_{n \geq 1} (K_n(t_{tot}) - K_n(t)) f_n'(z(t)) \right]^2 dz.$$

(8b)

The phase variance from the phase diffusion is [7]

$$\langle \phi_D(t_{tot})^2 \rangle = 2 \int_0^{t_{tot}} D_\phi(t) dt.$$

(9)

The second term, $\phi_{float}(t_{tot})$ in Eqs. (6a-b), can be rewritten as

$$\phi_{float}(t_{tot}) = \sum_{j=1}^{m} \left[ \sum_{n \geq 1} \left( K_n(t_{tot}) - K_n(t_j) \right) f_n''\left(z(t_j)\right) \right] D \Delta t_j,$$

(10)

where $D = \frac{(\Delta z_j)^2}{2\Delta t_j}$ is used. We can define a float velocity as

$$v_{float}(t) = \sum_{n \geq 1} (K(t_{tot}) - K(t)) f_n''(z(t)) D. \tag{11}$$

Note $D$ exists both in $D_\phi(t)$ and $v_{float}(t)$. In $D_\phi(t)$, $D$ comes from $\Delta z_j$, which is the fundamental diffusion parameter determining the random walk. While $D$ in $v_{float}(t)$ comes from $(\Delta z_j)^2$, which is just a positive constant to modify $\sum_{j=1}^{m} \left[ \sum_{n \geq 1} (K_n(t_{tot}) - K_n(t_j)) f_n''(z(t_j)) \right]$. If $f_n''(z(t))$ depends on $z(t)$, the $\sum_{j=1}^{m} \left[ \sum_{n \geq 1} (K_n(t_{tot}) - K_n(t_j)) f_n''(z(t_j)) \right]$ will be treated as a random walk process, demonstrated in the cubic field in section 2.2.2.

However, whether the second term is a drift motion or a diffusion, we can still define

$$\phi_{float}(t_{tot}) = \int_0^{t_{tot}} v_{drift}(t) dt. \tag{12}$$

When $f_n''(z(t))$ is an even function or $z_0^2 \gg 2Dt$,

$$\langle \phi_{float}(t_{tot}) \rangle = \int_0^{t_{tot}} (K_n(t_{tot}) - K_n(t)) \langle f_n''(z(t)) \rangle D dt, \tag{13}$$

which can be either positive or negative. While, when $f_n''(z(t))$ is an odd function and $z_0^2 < 2Dt$,

$$\langle \phi_{float}(t_{tot}) \rangle_\pm = c_{n,\pm,z_0} \int_0^{t_{tot}} (K_n(t_{tot}) - K_n(t)) \langle |f_n''(z(t))| \rangle D dt, \tag{14}$$

where $c_{n,\pm,z_0}$ is introduced because results from that in odd $n$, some particles' diffusion paths can cover both positive and negative regions when $z_0$ is near the origin, which reduces the individual integral and the total average. $c_{n,\pm,z_0}$ could be positive or negative depending on the pulse sequence and $z_0$. However, $|c(t)|$ should be near 1 as the $z(t)$ of a particle from different moments are correlated. The $\langle \phi_{float}(t_{tot}) \rangle$ and $\langle \phi_{float}(t_{tot}) \rangle_\pm$ can help us to understand better the phase change due to the second term in Eqs. (6a-b).

The phase variance, $\langle (\phi_{float}(t_{tot}))^2 \rangle$, can be calculated based on the effective phase diffusion or drift motion. Based on the effective phase diffusion,

$$\langle (\phi_{float}(t_{tot}))^2 \rangle = \langle \left( \int_0^{t_{tot}} (K_n(t_{tot}) - K_n(t)) f_n''(z(t)) D dt \right)^2 \rangle, \text{ if } f_n''(z(t)) \text{ depends on } z(t), \tag{15a}$$

while based on a drift evolution,

$$\langle (\phi_{float}(t_{tot}))^2 \rangle = (\langle \phi_{float}(t_{tot}) \rangle)^2, \text{ if } f_n''(z(t)) \text{ does not depend on } z(t). \tag{15b}$$

Eqs. (13, 15b) will be used to calculate the $(\langle \phi_{float}(t_{tot}) \rangle)^2$ for a drift evolution, and Eq. (15a) will be calculated based on the effective phase diffusion method.

From the third term in Eq. (6a),

$$\phi_{shift,z_0}(t_{tot}) = \sum_{n \geq 1} K_n(t_{tot}) f_n(z_0). \tag{16}$$

The general expressions for phase diffusion under a random nonlinear gradient field are eventually obtained based on the effective phase diffusion method.

If $B_g(z) = \sum_{n \geq 1} g_n(t) \cdot z^n$, namely, $f_n(z) = z^n$, which is a nonlinear gradient field with multiple integer orders whose $f_n'(z(t))$ and $f_n''(z(t))$ are

$$\begin{cases} f_n'(z(t)) = [(z(t))^n]' = n(z(t))^{n-1}, \\ f_n''(z(t)) = [(z(t))^n]'' = n(n-1)(z(t))^{n-2}. \end{cases} \tag{17}$$

Currently, the frequently encountered gradient field is the integer order gradient field. However, it may be desirable to develop the fraction-order gradient field, which could provide different features for gradient experiments. If $B_g(z) = \sum_{n\geq 1} g_n(t) \cdot z^{\frac{1}{n}}$, namely, $f_n(z) = z^{\frac{1}{n}}$, which is a fraction-order gradient field whose $f_n'(z(t))$ and $f_n''(z(t))$ are

$$\begin{cases} f_n'(z(t)) = \left[|z(t)|^{\frac{1}{n}}\right]' = \frac{z(t)(|z(t)|)^{\frac{1}{n}-2}}{n}, \\ f_n''(z(t)) = \left[|z(t)|^{\frac{1}{n}}\right]'' = -\frac{(n-1)(|z(t)|)^{\frac{1}{n}-2}}{n^2}. \end{cases} \qquad (18)$$

These $f_n'(z(t))$ and $f_n''(z(t))$ can be substituted into Eqs. (***) to calculate the phase evolution results. From the derivation, $g_n f_n(z)$ can be a random gradient function determined by real applications.

For simplicity, we will focus on the phase diffusion under a single-order nonlinear gradient field, $B(z,t) = B_0 + g_n(t) \cdot z^n$, where $n$ is an integer in the rest of this paper. An even and an odd order gradient field, the parabolic gradient field ($n = 2$), and an even order gradient field, the cubic gradient field ($n = 3$), will be used as examples for the calculation in this paper. The parabolic field has been studied in many literatures [10,12,13]. However, as mentioned in the introduction, the reported results have some apparent discrepancies. The results will help us understand the phase evolution under the nonlinear field, which could help develop advanced NMR and MRI experimental techniques based on the nonlinear field.

2.2 phase diffusion under $B(z,t) = B_0 + g_n(t) \cdot z^n$

The results in the previous section can be applied to single-order nonlinear gradient field $B(z,t) = B_0 + B_g(z)$, $B_g(z) = g_n(t) \cdot z^n$. From Eqs. (***), we have the following:

$$\phi_{shift,z_0}(t_{tot}) = K_n(t_{tot})(z_0)^n, \qquad (19)$$

$$D_\phi(t) = \langle\{(K_n(t_{tot}) - K_n(t))[(z(t))^n]'\}^2\rangle D = n^2(K_n(t_{tot}) - K_n(t))^2 \langle(z(t))^{2n-2}\rangle D, \qquad (20)$$

$$v_{float}(t) = (K(t_{tot}) - K(t))[(z(t))^n]'' D = (K_n(t_{tot}) - K_n(t))n(n-1)(z(t))^{n-2} D. \qquad (21)$$

i. When $n$ is even, $[(z(t))^n]''$ is even function, or $z_0^2 \gg 2Dt$, from Eqs. (12,21), we have

$$\langle\phi_{float}(t_{tot})\rangle = \int_0^{t_{tot}} v_{float}(t)dt = \int_0^{t_{tot}}(K_n(t_{tot}) - K_n(t))n(n-1)\langle(z(t))^{n-2}\rangle Ddt. \qquad (22)$$

ii. When $n$ is odd, $[(z(t))^n]''$ is an odd function and $z_0^2 < 2Dt$, from Eq. (12,21), we have

$$\langle\phi_{float}(t_{tot})\rangle_\pm = c_{n,\pm,z_0} \int_0^{t_{tot}}(K_n(t_{tot}) - K_n(t))n(n-1)\langle(z(t))^{n-2}\rangle Ddt. \qquad (23)$$

The total phase variance can be calculated based on:

$$\langle[\phi - \phi_{shift,z_0}(t_{tot})]^2\rangle = \langle(\phi_{float}(t_{tot}) + \phi_{diff}(t_{tot}))^2\rangle = \langle(\phi_{float}(t_{tot}))^2\rangle + \langle(\phi_D(t_{tot}))^2\rangle +$$
$$2\langle\phi_{float}(t_{tot})\rangle\langle(\phi_D(t_{tot}))\rangle \approx \langle(\phi_{float}(t_{tot}))^2\rangle + \langle(\phi_D(t_{tot}))^2\rangle. \qquad (24)$$

The NMR signal is the average magnetization for spin starting from $z_0$. When $n \neq 2$, $\phi_{float}(t_{tot})$ follows a distribution. If we assume that $\phi - \phi_{shift,z_0}(t_{tot})$ follows a Gaussian distribution, we have

$$S(t_{tot}) = exp\left(i\phi_{shift,z_0}(t_{tot})\right)|S(t_{tot})|, \qquad (25a)$$

where

$$|S(t_{tot})| = exp\left\{\frac{\left[\langle(\phi_{float}(t_{tot}))^2\rangle + \langle(\phi_D(t_{tot}))^2\rangle\right]}{2}\right\}, n \neq 2, \text{Gaussian phase distribution.} \qquad (25b)$$

When $n = 2$, the $\phi_{float}(t_{tot})$ does not follow a distribution. If we assume that $\phi_D(t_{tot})$ follows a

Gaussian distribution, we have

$$S(t_{tot}) \approx exp\{i[\phi_{float}(t_{tot}) + \phi_{shift,z_0}(t_{tot})]\}|S(t_{tot})|, \qquad (26a)$$

where

$$|S(t_{tot})| = exp\left[-\frac{\langle(\phi_D(t_{tot}))^2\rangle}{2}\right], n = 2, \text{Gaussian phase distribution.} \qquad (26b)$$

However, the correlation between the coefficients of the individual jump steps of the phase diffusion $\phi_{float}(t_{tot})$ and $\phi_D(t_{tot})$ may make the diffusion deviate from Gaussian diffusion. Other types of distributions, such as Lorentzian distribution or long-tailed phase distribution, could be assumed. The linewidth of the distribution could be

$$\Upsilon(t_{tot}) = \begin{cases} \frac{1}{\pi}\sqrt{\frac{[\langle(\phi_{float}(t_{tot}))^2\rangle + \langle(\phi_D(t_{tot}))^2\rangle]}{2}}, n \neq 2, \\ \frac{1}{\pi}\sqrt{\frac{\langle(\phi_D(t_{tot}))^2\rangle}{2}}, n = 2. \end{cases} \qquad (27)$$

And the amplitude of the signal attenuation $|S(t_{tot})|$ in Eqs. (22a) and (22b) will be replaced as

$$|S(t_{tot})| = \begin{cases} exp(-\Upsilon(t_{tot})), \text{Lorentzian phase distribution}, \\ E_\alpha(-\Upsilon(t_{tot})), \text{long} - \text{tailed fractional phase distribution}. \end{cases} \qquad (28)$$

where $E_\alpha(-\Upsilon(t_{tot}))$ is Mittag-Leffler type attunation. From our simulation, the signal attenuation for both parabolic field and cubic fields at the short time is Gaussian, then it changes to Lorentzian at the intermediate time; and at both intermediate time and longtime it obey Mittag-Leffler type attenuation. In Equation (27), $\sqrt{\frac{[\langle(\phi_{float}(t_{tot}))^2\rangle + \langle(\phi_D(t_{tot}))^2\rangle]}{2}}$ or $\sqrt{\frac{\langle(\phi_D(t_{tot}))^2\rangle}{2}}$ is used to obtain the $\Upsilon(t_{tot})$, which is based on the following considerations: First, the width of the half maximum is $\Upsilon(t_{tot})$ in Lorentzian distribution, while in Gaussian distribution, it is proportional to $\sqrt{\frac{[\langle(\phi_{float}(t_{tot}))^2\rangle + \langle(\phi_D(t_{tot}))^2\rangle]}{2}}$ or $\sqrt{\frac{\langle(\phi_D(t_{tot}))^2\rangle}{2}}$, which is tried in the simulations and found it provides a signal attenuation that agrees with the simulations. Additionally, the diffusion here is somewhat like the diffusion along a curvilinear path, which gives a MLF type PFG signal attenuation based on phase variance, $\sqrt{\frac{\langle(\phi_D(t_{tot}))^2\rangle}{2}}$ [24]. Similar phase variance dependence may be possible, although the situation is different in the phase diffusion here, where the diffusion coefficient distribution is affected by the curvilinear path. With more effort, theoretical signal attenuation may possibly be derived based on the curvilinear path-dependent diffusion coefficient.

We calculate the phase diffusion under three different pulses: a. $K_n(t_{tot}) \neq 0$, diffusion under $\frac{\pi}{2} - \delta$ r.f. pulse with a steady gradient field; b. $K_n(t_{tot}) = 0$, diffusion under $\frac{\pi}{2} - \delta - \pi - \delta$ r.f. pulse with a steady gradient field (PGSE or PGSTE, $\Delta = \delta$); c. $K_n(t_{tot}) = 0$, diffusion under pulsed gradient field (PGSE or PGSTE, $\Delta \geq \delta$). The $K_n(t_{tot}) - K_n(t)$, or $K_n(t)$ values are listed in Table 1.

**Table 1.** Three different field gradient pulses.

| |
|---|
| $K_n(t_{tot}) \neq 0$, diffusion under $\frac{\pi}{2} - \delta$ r.f. pulse with a steady gradient field, $t_{tot} = \delta$ |
| $K_n(t_{tot}) - K_n(t) = \gamma g_n \delta - \gamma g_n t$ |
| $K(t_{tot}) = 0$, diffusion under $\frac{\pi}{2} - \delta - \pi - \delta$ r.f. pulse with a steady gradient field (PGSE or PGSTE, $\Delta = \delta$), $t_{tot} = 2\delta$ |
| $K_n(t) = \begin{cases} -\gamma g_n t, 0 \leq t \leq \delta, \\ -\gamma g_n(2\delta - t), \delta \leq t \leq 2\delta. \end{cases}$ |
| $K(t_{tot}) = 0$, diffusion under pulsed gradient field (PGSE or PGSTE, $\Delta \geq \delta$), $t_{tot} = \Delta + \delta$ |
| $K_n(t) = \begin{cases} -\gamma g_n t, 0 \leq t \leq \delta, \\ -\gamma g_n \delta, \delta \leq t \leq \Delta, \\ -\gamma g_n(\Delta + \delta - t), \Delta \leq t \leq \Delta + \delta. \end{cases}$ |

One even order field, the parabolic field, $n = 2$, and one odd order field, the cubic field, $n = 3$, will be used as examples for the theoretical calculations. The calculation is shown in the following:

2.2.1    Parabolic field, $n = 2$, an even gradient field, $f_n''(z(t))$ does not depend on $z(t)$

Based on Eqs. (15-18), we get

$$\phi_{shift,z_0}(t_{tot}) = K_2(t_{tot})z_0^2. \tag{29}$$

$$D_\phi(t) = 4(K_2(t_{tot}) - K_2(t))^2 \langle (z(t))^2 \rangle D$$

$$\xrightarrow{\langle (z(t))^2 \rangle = z_0^2 + 2Dt} D_\phi(t) = 4(K_2(t_{tot}) - K_2(t))^2 (z_0^2 + 2Dt)D, \tag{30}$$

$$v_{float}(t) = \langle [K(t_{tot}) - K(t)] \cdot [(z(t))^2]'' \rangle D = 2[K(t_{tot}) - K(t)]D. \tag{31}$$

Because $[(z(t))^2]'' = 2$, which is a constant, it is not necessary to consider the distribution for $v_{float}(t)$. These Eqs. (20-22) can be substituted into Eqs. (12), (***) to calculate the phase variance. For $K(t_{tot}) = 0$, diffusion under pulsed gradient field (PGSE or PGST, $\Delta \geq \delta$), $t_{tot} = \Delta + \delta$, the calculated

$$\phi_{float}(t_{tot}) = 2\gamma g_2 D \delta \Delta, \tag{32}$$

and

$$\langle (\phi_D(t_{tot}))^2 \rangle = 8(\gamma g_2)^2 D^2 \left\{ \Delta^2 \delta^2 - \frac{\delta^4}{2} + \frac{1}{6}[(\Delta + \delta)^4 - 6(\Delta + \delta)^2 \Delta^2 + 8(\Delta + \delta)\Delta^3 - 3\Delta^4] \right\} + 8(z_0)^2(\gamma g_2 \delta)^2 D \left( \Delta - \frac{1}{3}\delta \right), \tag{33a}$$

Under short gradient pulse (SGP) approximation,

$$\langle (\phi_D(t_{tot}))^2 \rangle_{SGP} \approx 8(\gamma g_2)^2 D^2 \Delta^2 \delta^2 + 8(z_0)^2(\gamma g_2 \delta)^2 D\Delta. \tag{33b}$$

For $z_0 = 0$,

$$\langle (\phi_D(t_{tot}))^2 \rangle = 8(\gamma g_2)^2 D^2 \left\{ \Delta^2 \delta^2 - \frac{\delta^4}{2} + \frac{1}{6}[(\Delta + \delta)^4 - 6(\Delta + \delta)^2 \Delta^2 + 8(\Delta + \delta)\Delta^3 - 3\Delta^4] \right\}. \tag{34a}$$

and

$$\langle (\phi_D(t_{tot}))^2 \rangle_{SGP} \approx 8(\gamma g_2)^2 D^2 \Delta^2 \delta^2. \tag{34b}$$

**Table. 2**. Effective phase diffusion under parabolic gradient field, $n = 2$.

| $K_2(t_{tot}) \neq 0$, diffusion under $\frac{\pi}{2} - \delta$ r.f. pulse with a steady gradient field, $t_{tot} = \delta$ | | |
|---|---|---|
| | $z_0 \neq 0$ | $z_0 = 0$ |
| $\phi_{float}(t_{tot})$ | $\gamma g_2 D \delta^2$ | |
| $\langle(\phi_D(t_{tot}))^2\rangle$ | $\frac{8}{3}(z_0)^2(\gamma g_2)^2 D \delta^3 + \frac{4}{3}(\gamma g_2)^2 D^2 \delta^4$ | $\frac{4}{3}(\gamma g_2)^2 D^2 \delta^4$ |
| $\langle(\phi - (z_0)^2 \gamma g_2 \delta)^2\rangle$ | $(\gamma g_2)^2 D^2 \delta^4 + \frac{8}{3}(z_0)^2(\gamma g_2)^2 D \delta^3 + \frac{4}{3}(\gamma g_2)^2 D^2 \delta^4$ | $(\gamma g_2)^2 D^2 \delta^4 + \frac{4}{3}(\gamma g_2)^2 D^2 \delta^{4**}$ replicates the result in Ref. [12] |
| $S(t_{tot})$ | $exp\{i[\gamma g_2 D \delta^2 + (z_0)^2 \gamma g_2 \delta]\} exp\left[-\langle(\phi_D(t_{tot}))^2\rangle/2\right]$ replicates Eq. (3.9) in Ref. [10] | $exp\{i\gamma g_2 D \delta^2\} exp\left[-\frac{2}{3}(\gamma g_2)^2 D^2 \delta^4\right]$ |
| $K_2(t_{tot}) = 0$, diffusion under $\frac{\pi}{2} - \delta - \pi - \delta$ r.f. pulse with a steady gradient field ($\Delta = \delta$), $t_{tot} = 2\delta$ | | |
| | $z_0 \neq 0$ | $z_0 = 0$ |
| $\phi_{float}(t_{tot})$ | $-2\gamma g_2 D \delta^2$ | |
| $\langle(\phi_D(t_{tot}))^2\rangle$ | $\frac{16}{3}(z_0)^2(\gamma g_2)^2 D \delta^3 + \frac{32}{3}(\gamma g_2)^2 D^2 \delta^4$ | $\frac{32}{3}(\gamma g_2)^2 D^2 \delta^4$ |
| $\langle\phi^2\rangle$ | $4(\gamma g_2)^2 D^2 \delta^4 + \frac{16}{3}(z_0)^2(\gamma g_2)^2 D \delta^3 + \frac{32}{3}(\gamma g_2)^2 D^2 \delta^4$ | $4(\gamma g_2)^2 D^2 \delta^4 + \frac{32}{3}(\gamma g_2)^2 D^2 \delta^4$ agree with Ref. [12] |
| $S(t_{tot})$ | $exp\{i2\gamma g_2 D \delta^2\} exp\left[-\langle(\phi_D(t_{tot}))^2\rangle/2\right]$ | $exp\{i2\gamma g_2 D \delta^2\} exp\left[-\frac{16}{3}(\gamma g_2)^2 D^2 \delta^4\right]$, Ref. [12] reported $exp\left[-\frac{22}{3}(\gamma g_2)^2 D^2 \delta^4\right] = exp[-\langle\phi^2\rangle/2]$, Ref. [10] $]exp\left[-\frac{2}{3}(\gamma g_2)^2 D^2 \delta^3\right]$. |
| $K_2(t_{tot}) = 0$, diffusion under pulsed gradient field (PGSE or PGST, $\Delta \geq \delta$), $t_{tot} = \Delta + \delta$ | | |
| | $z_0 \neq 0$ | $z_0 = 0$ |
| $\phi_{float}(t_{tot})$ | $-2\gamma g_2 D \delta \Delta$ | |
| $\langle(\phi_D(t_{tot}))^2\rangle$ | $Eq.(32a)$ | $Eq.(32b)$ |
| $\langle\phi^2\rangle$ | $4(\gamma g_2)^2 D^2 \delta^2 \Delta^2 + \langle(\phi_D(t_{tot}))^2\rangle$ | $4(\gamma g_2)^2 D^2 \delta^2 \Delta^2 + \langle(\phi_D(t_{tot}))^2\rangle$ *Ref. $6(\gamma g_2)^2 D^2 \delta^2 \Delta \left(\Delta - \frac{2}{3}\delta\right)$ |
| $S(t_{tot})$ | $exp\{i2\gamma g_2 D \delta \Delta\}|S(t_{tot})|$, Eqs. (33a), (33b) and (35) | $,exp\{i2\gamma g_2 D \delta \Delta\}|S(t_{tot})|$ Eqs. (34a), (34b) and (35) |

The signal is

$$S(t_{tot}) = exp\{-i2\gamma g_2 D \delta \Delta\}|S(t_{tot})|, \qquad (35)$$

where $|S(t_{tot})|$ is described by Eqs. (26b) for Gaussian phase distribution, and (28) for Lorentzian and long-tailed phase distributions.

The calculated $\phi_D(t_{tot})$, $\phi_{float}(t_{tot})$, and $\phi_{shift,z_0}(t_{tot})$ with various conditions, and the corresponding magnetizations for NMR or MRI gradient experiments affected by the parabolic field are listed in Table 2. The results for $\Delta \geq \delta$ can be reduced to the result for $\Delta = \delta$ presented in Table 2.

2.2.2 Cubic field, $n = 3$, an odd gradient field, $f_n''(z(t))$ depends on $z(t)$

Based on Eqs. (****), we get

$$\phi_{shift,z_0}(t_{tot}) = K_3(t_{tot})z_0^3. \tag{36}$$

$$D_\phi(t) = 9(K_3(t_{tot}) - K_3(t))^2 \langle (z(t))^4 \rangle D \xrightarrow{(z(t))^4 = 12Dtz_0^2 + 12(Dt)^2 + z_0^4} D_\phi(t) = 9(K_3(t_{tot}) - K_3(t))^2 (12Dtz_0^2 + 12(Dt)^2 + z_0^4)D. \tag{37a}$$

For $z_0 = 0$ and $z_0 \gg 2Dt$, $D_\phi(t)$ can be approximated as

$$D_\phi(t) = \begin{cases} 108[K_3(t_{tot}) - K_3(t_j)]^2 D^3 t^2, z_0 = 0 \\ 9 \times (z_0)^4 [K_3(t_{tot}) - K_3(t_j)]^2 D, z_0 \gg 2Dt. \end{cases} \tag{37b}$$

It is not difficult to calculate $\langle (\phi_D(t_{tot}))^2 \rangle$ based on Eq. (9). Note $\langle (z(t) - Z_0)^4 \rangle \neq ((z(t) - z_0)^2)^2$.

$$\langle (\phi_D(t_{tot}))^2 \rangle_{z_0=0} = 2 \int_0^{t_{tot}} 9(K_3(t_{tot}) - K_3(t))^2 (12Dtz_0^2 + 12(Dt)^2 + z_0^4)Ddt. \tag{38}$$

$\left[ (z(t))^3 \right]'' = 6z(t)$ can be substituted into Eq. (10) to give

$$\phi_{float}(t_{tot}) = \sum_{j=1}^m (K_3(t_{tot}) - K_3(t_j)) 6z(t_j) D\Delta t_j = \sum_{j=1}^m 6[K_3(t_{tot}) - K_3(t)](z(t) - z_0)D\Delta t_j + \sum_{j=1}^m 6[K_3(t_{tot}) - K_3(t)]z_0 D\Delta t_j = \phi_{float,d}(t_{tot}) + \phi_{float,s}(t_{tot}), \tag{39a}$$

where

$$\phi_{float,d}(t_{tot}) = \sum_{j=1}^m 6[K_3(t_{tot}) - K_3(t)](z(t) - z_0)D\Delta t_j, \tag{39b}$$

and

$$\phi_{float,s}(t_{tot}) = \sum_{j=1}^m 6[K_3(t_{tot}) - K_3(t)]z_0 D\Delta t_j. \tag{39c}$$

By setting $Z(t) = z(t) - z_0$, $\gamma' = 6D$, $g'(t) = [K_3(t_{tot}) - K_3(t)]$,

$$\phi_{float,d}(t_{tot}) = \int_0^{t_{tot}} \gamma' g'(t) Z(t) dt. \tag{39d}$$

Because $Z(t)$ is the diffusing distance from the starting point in the new relative reference, the phase evolution of $\int_0^{t_{tot}} \gamma' g'(t) Z(t) dt$ can be calculated based on the effective phase diffusion method, where $\int_0^{t_{tot}} \gamma g(t) Z(t) dt$ has been obtained or based on Eqs. (7,9) presented in this paper. Note in the calculation, $K'_3(t_{tot}) \neq 0$ for $K'_3(t) = \int_0^t \gamma' g'(t') dt'$.

The calculated results are listed in Table 3.

**Table. 3**. Effective phase diffusion under parabolic gradient field, $n = 3$.

| $K_3(t_{tot}) \neq 0$, diffusion under $\frac{\pi}{2} - \delta$ r.f. pulse with a steady gradient field, $t_{tot} = \delta$ | | |
|---|---|---|
| | $z_0 \geq 0$ | $z_0 = 0$ |
| $\langle(\phi_{float}(t_{tot}))^2\rangle$ | $(3\gamma g_3 z_0 D\delta^2)^2 + \frac{18}{5}(\gamma g_3)^2 D^3\delta^5$ $\langle(\phi_{float,d}(t_{tot}))^2\rangle = \frac{18}{5}(\gamma g_3)^2 D^3\delta^5$ $\langle(\phi_{float,s}(t_{tot}))^2\rangle = (3\gamma g_3 z_0 D\delta^2)^2$ | $\frac{18}{5}(\gamma g_3)^2 D^3\delta^5$ |
| $\langle(\phi_D(t_{tot}))^2\rangle$ | $6(z_0)^4(\gamma g_3)^2 D\delta^3 + 18(z_0)^2(\gamma g_3)^2 D^2\delta^4 + \frac{36}{5}(\gamma g_3)^2 D^3\delta^5$ | |
| $\langle(\phi - (z_0)^3\gamma g_3\delta)^2\rangle$ | $\langle(\phi_{drift}(t_{tot}))^2\rangle + \langle(\phi_D(t_{tot}))^2\rangle$ | |
| $S(t_{tot})$ | $exp[i(z_0)^3 K_3(t_{tot})] \, |S(t_{tot})|$ | $exp\left[-\frac{54}{5}(\gamma g_3)^2 D^3\delta^5\right]$ |
| $K_3(t_{tot}) = 0$, diffusion under $\frac{\pi}{2} - \delta - \pi - \delta$ r.f. pulse with a steady gradient field ($\Delta = \delta$), $t_{tot} = 2\delta$ | | |
| | $z_0 \geq 0$ | $z_0 = 0$ |
| $\langle(\phi_{float}(t_{tot}))^2\rangle$ | $(\phi_{float,d}(t_{tot}))^2 + \langle(\phi_{float,s}(t_{tot}))^2\rangle$ $\langle(\phi_{float,d}(t_{tot}))^2\rangle = \frac{276}{5}(\gamma g_3)^2 D^3\delta^5$ $\langle(\phi_{float,s}(t_{tot}))^2\rangle = (6\gamma g_3 z_0 D\delta^2)^2$ | $\frac{276}{5}(\gamma g_3)^2 D^3\delta^5$ |
| $\langle(\phi_D(t_{tot}))^2\rangle$ | $\left[\frac{792}{5}D^3\delta^3 + 144(z_0)^2 D^2\delta^2 + 12(z_0)^4 D\delta\right](\gamma g_3\delta)^2$ | $\frac{792}{5}(\gamma g_3)^2 D^3\delta^5$ |
| $\langle\phi^2\rangle$ | $\langle(\phi_{drift}(t_{tot}))^2\rangle + \langle(\phi_D(t_{tot}))^2\rangle$ | |
| $S(t_{tot})$ | $|S(t_{tot})|$ | |
| $K_3(t_{tot}) = 0$, diffusion under pulsed gradient field (PGSE or PGST, $\Delta \geq \delta$), $t_{tot} = \Delta + \delta$ | | |
| | $z_0 \geq 0$ | $z_0 = 0$ |
| $\langle(\phi_{float}(t_{tot}))^2\rangle$ | $(\phi_{float,d}(t_{tot}))^2 + \langle(\phi_{float,s}(t_{tot}))^2\rangle$ $\langle(\phi_{float,d}(t_{tot}))^2\rangle = 2(\gamma g_3)^2 D^3\left(\frac{3}{5}\delta^5 - 3\Delta\delta^4 + 18\Delta^2\delta^3 + 12\Delta^3\delta^2\right)$ $\langle(\phi_{float,s}(t_{tot}))^2\rangle = (6\gamma g_3 z_0 D\Delta\delta)^2$ | $2(\gamma g_3)^2 D^3\left(\frac{3}{5}\delta^5 - 3\Delta\delta^4 + 18\Delta^2\delta^3 + 12\Delta^3\delta^2\right)$ |
| $\langle(\phi_D(t_{tot}))^2\rangle$ | $\left[-\frac{108}{5}D^3\delta^3 + 36D^3\Delta\delta^2 + 72\Delta^2 D^3\delta + 72\Delta^3 D^3 - 36(z_0)^2 D^2\delta^2 + 72(z_0)^2\Delta D^2\delta - 6(z_0)^4 D\delta + 108(z_0)^2\Delta^2 D^2 + 18(z_0)^4 D\Delta\right](\gamma g_3\delta)^2$ | $\left[-\frac{108}{5}D^3\delta^3 + 36D^3\Delta\delta^2 + 72\Delta^2 D^2\delta + 72\Delta^3 D^3\right](\gamma g_3\delta)^2$ |
| $\langle\phi^2\rangle$ | $\langle(\phi_{float}(t_{tot}))^2\rangle + \langle(\phi_D(t_{tot}))^2\rangle$ | |
| $S(t_{tot})$ | $|S(t_{tot})|$ | |

The signal is

$$S(t_{tot}) = \begin{cases} exp[i(z_0)^3 K_3(t_{tot})]|S(t_{tot})|, K_3(t_{tot}) \neq 0, \\ |S(t_{tot})|, K_3(t_{tot}) = 0. \end{cases} \quad (40)$$

where $|S(t_{tot})|$ is described by Eqs. (25b) for Gaussian phase distribution, and (28) for Lorentzian and long-tailed phase distributions. From Table 3, we can obtain the total phase variance by the SGP approximation as

$$\langle(\phi(t_{tot}))^2\rangle_{SGP} \approx [72\Delta^3 D^3 + 72(z_0)^2 \Delta D^2 \delta - 6(z_0)^4 D\delta + 108(z_0)^2 \Delta^2 D^2 + 18(z_0)^4 D\Delta](\gamma g_3 \delta)^2$$
$$+ 24(\gamma g_3 \delta)^2 D^3 \Delta^3 + (6\gamma g_3 z_0 D\Delta\delta)^2. \quad (41)$$

which can be substituted into Eqs. (25a-b, 28) to obtain the corresponding SGP signal.

### 3. SIMULATION

Numerical simulation has been a convenient tool for verifying PFG diffusion theoretical results for linear gradient fields [25]. Here, one-dimensional discrete random walk simulations were performed to verify the theoretical results obtained in this paper. In the simulation, the jump length is $\epsilon$, and the jump waiting time is $\tau$, respectively, and real space diffusion constant $D = \frac{\epsilon^2}{2\tau}$. The accumulating spin phase associated with the diffusion path is recorded based on $\sum_{i=1}^{m} \gamma \Delta t_i g_n(t_i) \cdot z^n$. The PFG signal attenuation in the simulation can be obtained by averaging over all the walkers in the simulation [25].

The simulation counts the frequency of the particles appearing in each range of the coordinate in the phase space at the last record time of each random walk. The frequency is then divided by the width of the range to give the average frequency. The average frequency divided by the total number of random walks gives the probability distribution function. A similar strategy has been applied to obtain the real space probability density distribution in Ref. [26]. For each simulation, the total time span for each random walk is $4000\tau$, and 20 k repetitions of random walks are used.

### 4. RESULTS AND DISCUSSION

An effective phase diffusion method for describing nonlinear gradient field experiments has been developed in this paper. Compared to traditional methods, this method has obvious advantages: First, it can treat random shape gradient pulses with random order gradient fields, including integer and non-integer orders, even fraction orders. While some traditional methods, such as the Green function method [10] and the method in Ref. [12], currently only deal with the parabolic field, $n = 2$. Second, it reveals that three significantly different types of virtual phase evolutions, $\phi_D(t_{tot})$, $\phi_{float}(t_{tot})$, and $\phi_{shift,z_0}(t_{tot})$, exist in nonlinear gradient field experiments. For $n = 2$, the float phase $\phi_{float}(t_{tot})$ is a drift motion, which changes the final phase shift of the signal but does not affect its amplitude. While for $n \neq 2$, $\phi_{float}(t_{tot})$ is a random walk, affecting the total phase variance and, consequently, the signal amplitude. Third, the approximate total phase variances can be obtained, and their corresponding signal attenuations agree with the simulations. Fourth, this method can handle diffusion starting from any location, the origin, or other place. Theoretical treatment of the diffusion starting from the origin is more complicated than that from non-origin because $(z(t))^n$ cannot be approximated as $z_0^n$, which is a barrier that many traditional approximate theoretical methods cannot overcome. However, diffusion starting from or near the origin is an important part of the nonlinear gradient field phenomenon.

These three types of virtual phase evolutions: $\phi_D(t_{tot})$, $\phi_{float}(t_{tot})$, and $\phi_{shift,z_0}(t_{tot})$ are significantly different. $\phi_D(t_{tot})$ and $\phi_{float}(t_{tot})$ result from $f_n'(z(t))$ and $f_n''(z(t))$, the first and second derivatives of the gradient field, respectively. $\phi_{float}(t_{tot})$ could either be a drift motion or a phase diffusion. When $f_n''(z(t))$ is a constant, it is a drift motion, such as in the parabolic field case. However, it is a phase random walk when $f_n''(z(t))$ is a $z(t)$-dependent function, such as in the cubic field case. The $\phi_{shift,z_0}(t_{tot})$ is a $z_0$ dependent non-refocusing phase for $K(t_{tot}) \neq 0$, a pure phase shift in the final signal.

Recognizing and distinguishing the three types of phase evolutions is essential to determining the correct NMR signal expression. The reported literature does not have clear concepts of these three types of phase evolutions. In particular, the effect of $\phi_{float}(t_{tot})$ is missed or misplaced in NMR signal expression in traditional theoretical results, which can lead to the wrong signal phase or amplitude. The effects of $\phi_{float}(t_{tot})$ on the NMR signal are significant different between the drift motion and the phase random walk. In the drift motion, $\phi_{float}(t_{tot})$ affects the signal phase, as shown in Eqs. (26a-b), and it does not affect the signal attenuation amplitude. In contrast, the phase random walk contributes to the total phase variance that affects the signal attenuation amplitude, as shown in Eqs. (25a-b). Figure 1 shows that at small attenuation, for spin starting diffusion from the origin of the nonlinear gradient field, the signal attenuation amplitude in the parabolic field and cubic field based on the expressions listed in Tables 2 and 3 have good agreement with simulations. For the parabolic field, both the theoretical and the simulation results in this paper indicate that the drift phase does not affect the signal attenuation. While, Ref. [12] does not distinguish $\phi_{float}$ and $\phi_D$, and give $\exp\left(-\frac{7}{6}(\gamma g_2)^2 D^2 \delta^4\right)$, and $\exp\left(-\frac{22}{3}(\gamma g_2)^2 D^2 \delta^4\right)$ for signal attenuation in one-pulse and two-pulse experiments, respectively. These expressions overcount the signal attenuation and disagree with the simulations, as shown in Figure 1a. While the $\exp\left[-\frac{\langle(\phi_D(t_{tot}))^2\rangle}{2}\right]$, $\exp\left(-\frac{2}{3}(\gamma g_2)^2 D^2 \delta^4\right)$, and $\exp\left(-\frac{16}{3}(\gamma g_2)^2 D^2 \delta^4\right)$ obtained in this paper agrees with the simulation. The differences in the phase variances between the results in Ref. [12] and this paper are equal to $(\gamma g_2)^2 D^2 \delta^4$ and $2(\gamma g_2)^2 D^2 \delta^4$ for one and two pulse sequences, respectively, which exactly equal $\langle(\phi_{float}(t_{tot}))^2\rangle$. Therefore, the total phase variance in Ref. [12] should include both $\langle(\phi_{float}(t_{tot}))^2\rangle$ and $\langle(\phi_{float}(t_{tot}))^2\rangle$, which misplaces the drift phase and thus leads to overcounted signal attenuation. Ref. [10] gives $\exp\left(ig_2 D \delta^2 - \frac{2}{3}(\gamma g_2)^2 D^2 \delta^4\right)$ for NMR signal expression, the same as the result for one gradient field pulse sequence obtained in this paper. From the above discussion, it is clear that the discrepancy between Ref. [10] and [12] results from where the $\phi_{float}(t_{tot})$ is placed in the NMR signal expression, in the phase shift, or the attenuation. Except for the parabolic field, $\exp\left[-\frac{\langle(\phi_{float}(t_{tot}))^2\rangle + \langle(\phi_D(t_{tot}))^2\rangle}{2}\right]$ should be used as signal attenuation expression.

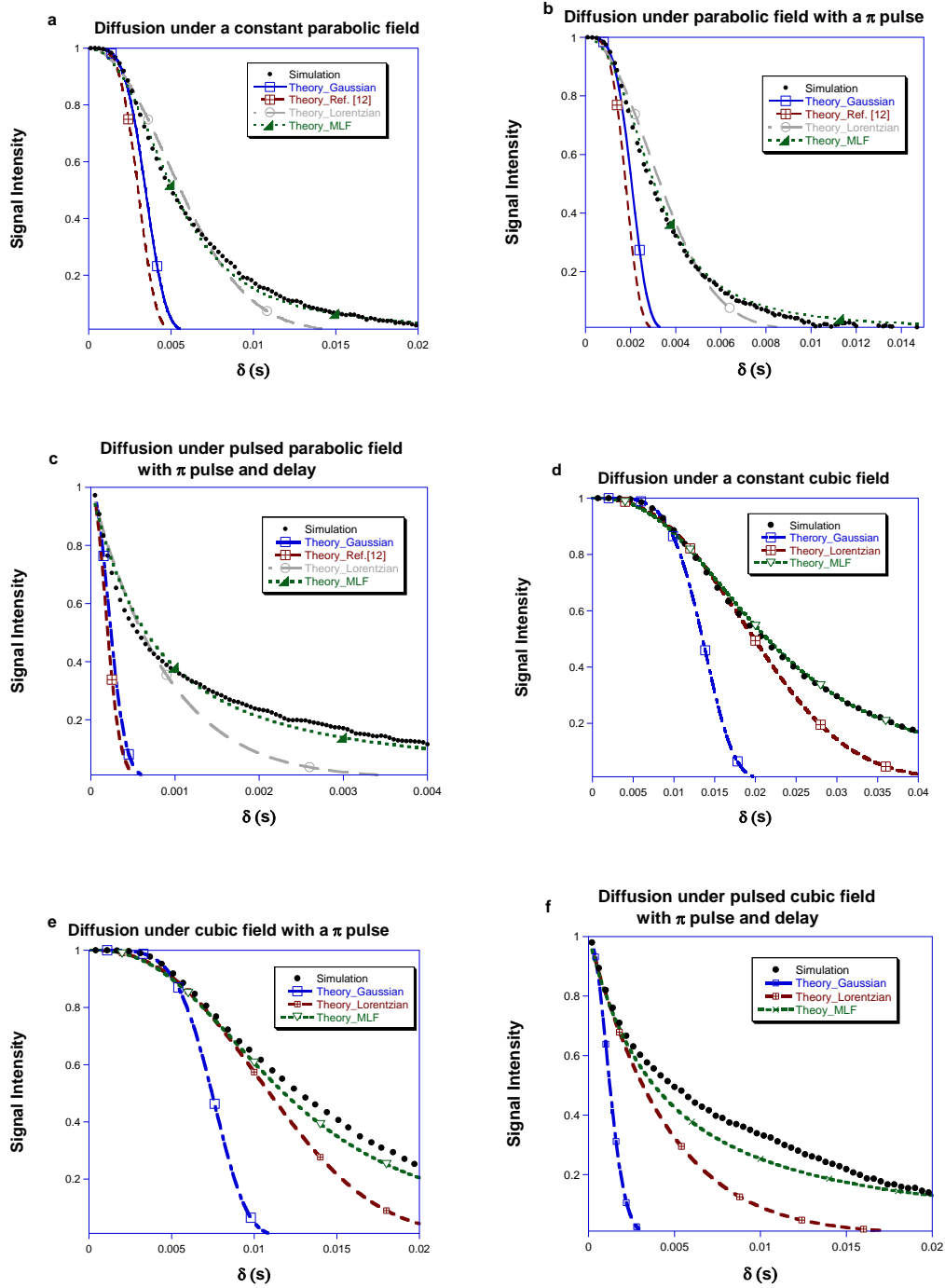

**Fig. 1** Comparison of theoretical NMR signal attenuations with random walk simulations. (a), (b) and (c) are for diffusion affected by the parabolic field, where $g_2$=16T/cm². (d), (e) and (f) are for diffusion affected by the cubic field, where $g_3$=704T/cm³. $\Delta - \delta = 20$ ms are used for (c) and (f), and $\gamma = 2.675 \times 10^8$ rad/s/T, $D = 2 \times 10^{-9}$ m²/s is used for (a-f). The signal attenuation obeys Gaussian attenuation for a short time, then changes to Lorentzian or MLF attenuations.

The $\phi_{float}(t_{tot})$ results in the oscillatory phase term in the parabolic field. Traditional methods employing the effective gradient [10,18] often deal with the first derivative of the field but are seldom aware of the effect of the second derivative of the magnetic field on the signal. The effect of $\phi_{float}(t_{tot})$ has appeared in Refs [10], but only in the case of diffusion under a constant gradient pulse, where an extra oscillatory phase term $ig_2Dt^2$ is presented, which is the same as $\phi_{float}(t_{tot})$ the $K_n(t_{tot}) \neq 0$ case in Table 2. However, Ref. [10] does not have similar oscillatory terms in the results of two pulses with $K(t_{tot}) = 0$. From table ***, for $K(t_{tot}) = 0$, the phase shifts from $\phi_{float}(t_{tot})$ are $-2\gamma g_2 D\delta\Delta$. The same result as $-2\gamma g_2 D\delta\Delta$ has also been obtained in Ref. [27] by averaging (summarizing) all possible phases along the diffusion paths. However, Ref. [28] only obtained the average phase shift for diffusion starting from the origin, which can be genderized to starting from random position as shown in Appendix A. The results clearly show that $\phi_{float}(t_{tot})$ are $-2\gamma g_2 D\delta\Delta$, which does not depend on the starting position $z_0$. $-2\gamma g_2 D\delta\Delta$ reduces to $-2\gamma g_2 D\delta^2$ when $\Delta = \delta$. It is essential to include $\phi_{float}(t_{tot})$ in analyzing the NMR signal. It is possible to use the $\phi_{float}(t_{tot})$ equaling $-2\gamma g_2 D\delta\Delta$ to measure the diffusion coefficient in the parabolic field for a short time. When time increases, the phase distribution is asymmetric as shown in Figure 2. The asymmetry should come from the correlation between the consecutive jumps; the phase jump length is $z(t)$ dependent. Along the spin diffusion path in real space, the long $z(t)$ follows long $z(t)$, which modifies the phase jump lengths. In the real NMR signal, the measure phase shift is not the average phase, but is the phase of the average magnetization vector, which depends on $arctan\frac{\langle sin(-\phi)\rangle}{\langle cos(-\phi)\rangle}$. Beside the float phase $\phi_{float}(t_{tot})$, the shift due to the asymmetry around the phase distribution peak affects the total phase shift for $S(t_{tot})$ in Eqs. (26a) and (35). However, an analytical expression for the shift due to both the float phase evolution and the asymmetry is not available now.

The phase distributions in even and odd gradient fields are different. Figure 2 shows that the phase distribution is asymmetric in the parabolic field. The left side of the distribution peak looks like Gaussian, but the right side looks like Lorentzian or long-tailed distribution. Meanwhile, the phase distribution in the cubic field is Lorentzian or long tailed. The phase shifts obtained in even-order gradient fields are all positive or negative and symmetric around the origin of the gradient field. However, the populations of phase distribution on different sides of the distribution peak are asymmetric. While the phase shifts under odd-order gradient fields are asymmetric, their population of phase distribution is symmetric around the origin.

The Lorentzian or long-tailed phase distribution, as shown in Figure 2, results in non-Gaussian signal attenuation. Unlike the linear gradient field, for the diffusion starting from the origin, the attenuation deviates from Gaussian when diffusion time is not short. As shown in Figure 1, the Gaussian attenuation only works for a very short period. After that period, it changes to Lorentzian and MLF attenuation [7]. The MLF are evaluated based on Pade approximation [28]. The Lorentzian and MLF attenuation is not from directly theoretical derivation but is proposed based on the simulation's phase distribution.

The Mittag-Leffler attenuation is very close to the Lorentzian attenuation at small attenuation. This phenomenon is reasonable because in a short time, $\Upsilon(t_{tot})$ is small, and the Mittag-Leffler function, $E_\alpha(-\Upsilon(t_{tot})) \approx \exp\left(-\frac{\Upsilon(t_{tot})}{\Gamma(1+\alpha)}\right)$. For a long time, the Mittag-Leffler function attenuates slower than the exponential function. Traditional methods often assume a Gaussian distribution, which has difficulty in interpreting the signal attenuation in a nonlinear field. In this paper, $E_\alpha(-\Upsilon(t_{tot}))$ rather than $E_\alpha(-\Gamma(1+\alpha)\Upsilon(t_{tot}))$ is used for the attenuation because $\Upsilon(t_{tot})$ rather than $\Gamma(1+\alpha)\Upsilon(t_{tot})$ is obtained from the phase variance. Currently, it is unsure what value the derivative order $\alpha$ should be used. From simulation, $\alpha$ equals 0.75 in Figures 1a-b, 0.5 in Figure 1c, and 0.3 in Figures 1d-f. The deviation from Gaussian should result from the correlation between the jump steps mentioned above. Additionally, in the even order gradient field, the asymmetry causes more complicated deviation. Currently, it is still a challenge to directly derive non-Gaussian phase distribution and the corresponding signal attenuations, such as Lorentzian and MLF attenuations. It still needs more research efforts.

Although the longtime phase distribution deviates from the Gaussian distribution, Eqs. (9,24, 39d) is still a good approximation for obtaining the phase variance because the phase diffusion could be viewed as a multiple-component diffusion. Each of the diffusions obeys the Gaussian diffusion. The phase diffusion coefficient is the average of the diffusion coefficients of all diffusion components. Thus, the Lorentzian or MLF attenuation signal attenuation based on the total variance still shows good agreement with the simulation.

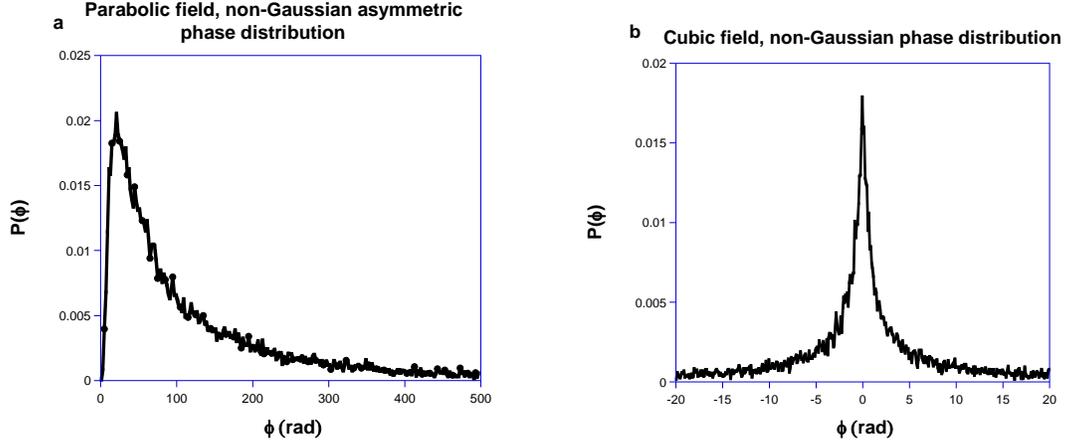

**Fig. 2** Phase distribution of spin diffusion under the parabolic and cubic fields from simulation: (a) asymmetric distribution in the parabolic field, $g_2$=16T/cm², (b) Lorentzian or long-tailed distribution in the cubic field, $g_3$=704T/cm³. Other parameters are $\Delta - \delta = 20$ ms, $\delta = 10$ ms, $\gamma = 2.675 \times 10^8$ rad/s/T, and $D = 2 \times 10^{-9}$ m²/s.

When $z_0 \gg 2Dt$, the signal attenuation is Gaussian, as shown in Figure 3. The Gaussian attenuation is reasonable because the jump length modifier $(z(t))^3 \approx z_0^n$, which is constant at each jump step, and the random walk can be approximated as a Gaussian random walk. As the measured NMR phase depends on $arctan\frac{\langle sin(-\phi)\rangle}{\langle cos(-\phi)\rangle}$, for spins starting diffusion far away from the origin, the effect of the float phase on the total phase shift still cannot be neglected for an even order gradient field, although its absolute value is smaller than that of the diffusion phase. The contribution to the total phase depends on how the final phase appears in the phase region between -$\pi$ and $\pi$, not the absolute phase values.

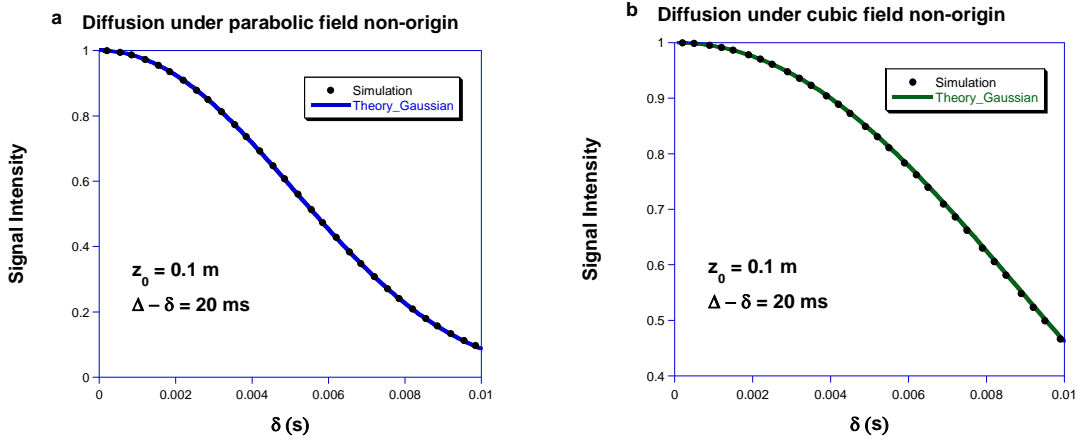

**Fig. 3** Signal attenuation under the parabolic and cubic fields for $z_0 \gg 2Dt$ from the simulation: (a) parabolic field, $g_2$=0.4T× $10^{-4}$/cm², (b) cubic field, $g_3$=1.5× $10^{-6}$T/cm³. Other parameters are $z_0 = 0.1$ m, $\Delta - \delta = 20$ ms, $\gamma = 2.675 \times 10^8$ rad/s/T, and $D = 2 \times 10^{-9}$ m²/s.

The nonlinear gradient field has certain advantages. Because $\langle (z(t))^{2(n-1)} \rangle = 2 \times \frac{1}{\sqrt{4\pi Dt}} \frac{(2n-3)!!}{2^n \left(\frac{1}{4Dt}\right)^4} \sqrt{\frac{\pi}{\frac{1}{4Dt}}} \propto (Dt)^{n-1}$, which can be substituted into Eqs. (20,9,) and eventually obtain $\langle \phi^2 \rangle \propto D^n t^{n+2}$. Therefore, for $n$ order nonlinear gradient field, the phase variance for the spins starting diffusion from the origin should be closely related to $\langle \phi^2 \rangle \propto D^n t^{n+2}$. Such a higher-exponent dependence could provide increased contrast factors for diffusion coefficient and time in NMR and MRI. These dependencies are different at the origin and non-origin positions. Selecting origin or non-origin positions should be based on the need for experiments.

In our results, the signal expression $S(t_{tot})$ describes spins starting diffusion from the same position $z_0$. In actual applications, this could be obtained by selective pulses. In typical experiments, the signal should come from diffusion starting from all places in the sample. The total signal will be $S_{whole\,sample}(t_{tot}) \propto \int_{z_a}^{z_b} S_z(t_{tot}) dz$, where $S_z(t_{tot})$ is the signal $S(t_{tot})$ presented in this paper.

The current paper focuses on providing general theoretical expressions for nonlinear fields. The parabolic and cubic fields are used as examples, and simulations are used to verify the theoretical results. The general expressions can be applied to calculate the signal attenuation for $z_0 \approx 2Dt$, which will not be pursued here. More efforts can be applied to use this method to investigate various situations such as higher-order gradient-field. Additionally, this method can be extended to handle fractional diffusion as well. In the practical application of the parabolic field, $x^2$, $y^2$, and $z^2$ could coexist [27], which can be calculated straightwardly by the method proposed in this paper. This method provides a broader view of phase evolution under the influence of a nonlinear gradient field, including various aspects: phase diffusion coefficient, phase variance, phase distribution, and signal attenuation. This method is versatile. The results can help develop advanced nonlinear gradient experimental techniques for NMR and MRI.

# APPENDIX: ACCUMULATED PHASE SHIFT FOR SPIN STARTING FROM RANDOM POSITION

In Ref. [27], the phase shift from the origin is obtained by averaging all possible phases along the diffusion paths, which is

$$\langle \phi \rangle = \int_0^{t_{tot}} \gamma g_2(t) \left[ \int_{-\infty}^{\infty} (z(t))^2 \frac{1}{\sqrt{4\pi Dt}} exp\left(-\frac{(z(t))^2}{4Dt}\right) \right] dt. \tag{A1}$$

which gives $-2\gamma g_2 D\delta\Delta$ for $K_2(t_{tot}) = 0$, diffusion under pulsed gradient field (PGSE or PGST, $\Delta \geq \delta$), $t_{tot} = \Delta + \delta$. Eq. (A1) can be generalized to all spins starting diffusion from a random position as

$$\langle \phi \rangle = \int_0^{t_{tot}} \gamma g_2(t) \left[ \int_{-\infty}^{\infty} (z(t))^2 \frac{1}{\sqrt{4\pi Dt}} exp\left(-\frac{(z(t)-z_0)^2}{4Dt}\right) \right] dt = \int_0^{t_{tot}} \gamma g_2(t) \left[ \int_{-\infty}^{\infty} ((z_0)^2 + 2Dt) \right] dt$$
$$\xrightarrow{K_2(t_{tot})=0} -2\gamma g_2 D\delta\Delta. \tag{A2}$$

Therefore, $\langle \phi \rangle$ is always $-2\gamma g_2 D\delta\Delta$ regardless of the starting position in the parabolic field, which agrees with $\phi_{float}(t_{tot})$ obtained in this paper. This generalization strategy can be applied to obtain average phase shifts affected by different order gradient fields. The average phase shift exists in all even-order gradient fields, which, however, is not the measured phase shift in NMR experiments. The measured phase shit is the phase of the average magnetization vector, depending on $arctan \frac{\langle sin(-\phi) \rangle}{\langle cos(-\phi) \rangle}$.